\documentstyle[prl,aps,twocolumn,epsfig]{revtex}
\begin{document}

\title{Tunneling Proximity Resonances:\ Interplay between 
Symmetry and Dissipation}

\author{Kristi Pance, Lorenza Viola, and S. Sridhar\thanks{Electronic 
addresses: kpance@sagar-3.physics.neu.edu; vlorenza@mit.edu; 
srinivas@neu.edu. } }

\address{Physics Department, Northeastern University, 360 Huntington Avenue, 
Boston, MA\ 02115}
\date{\today}

\maketitle

\begin{abstract}
We report the first observation of bound-state proximity resonances in
coupled dielectric resonators. The proximity resonances arise from the
combined action of symmetry and dissipation. We argue that the large ratio
between the widths is a distinctive signature of the multidimensional nature
of the system. Our experiments shed light on the properties of 2D tunneling
in the presence of a dissipative environment.
\end{abstract}

\vskip1.5pc

Tunneling and dissipation are ubiquitous phenomena in physics. A detailed
understanding of their combined action would be highly desirable given the
relevance of the problem for atomic physics, condensed matter physics,
chemistry and biology \cite{tunneling}. However, the incorporation of
dissipative effects is by no means trivial. Due to limitations of the
available analytical and computational methods, up-to-date descriptions are
still restricted to a few manageable cases, the prototype situation
involving a bistable potential in 1D \cite{leggett}.

In this Letter we report the observation of novel aspects of tunneling in 2D
potentials and its interplay with classical dissipation. In experiments
utilizing microwave dielectric resonators, we find that symmetry not only
plays a crucial role while shaping the eigenstates of the system, but also
influences the way they couple to the external environment acquiring a
finite width. In the observed resonance multiplets, we find that one of the
members is extremely sharp due to the symmetry of the configuration. The
large ratios of the observed widths appear to be a peculiar consequence of
the multidimensional nature of the system.

The experiments were carried out using $MgTi$ dielectric cylinders placed
between two parallel copper plates, 30 cm square, separated by a gap $l=6.38$
mm (Fig. 1). The disks had diameter $D=12.65$ mm and dielectric constant 
$\varepsilon _{r}=16$. After establishing input/output coupling to the {\sl 
near} field of the resonators by inserting coax lines terminated by loops,
measurements of the transmission amplitude as a function of the frequency
were performed using an HP8510B network analyzer.

The eigenvalue problem of a single dielectric resonator can be solved
analytically by regarding the system as a waveguide along the direction $z$
orthogonal to the plates \cite{Kajfeh88}. The entire field configuration can
be derived from the knowledge of the longitudinal components $\left\{
H_{z},E_{z}\right\} $ alone, that separately obey the Helmoltz equation:
\begin{equation}
(\nabla ^{2}+k^{2})\{H_{z}({\bf r}),E_{z}({\bf r})\}=0\,.
\end{equation}
Here, ${\bf r=(}\phi ,\rho ,z)$ in cylindrical coordinates and $k=\sqrt{
\varepsilon _{r}}\,(\omega /c)$ denotes the medium wave number for a mode at
frequency $\omega =2\pi f$ ($\varepsilon _{r}=1$ outside the dielectric).
For perfectly conducting walls, boundary conditions require $k_{z}$ $=p\pi
/l,$ $p$ integer. We have verified through explicit measurements of the
field profile that $p\geq 1$, and all modes are evanescent with a decay
constant close to the expected value $\kappa _{r}=\sqrt{k_{z}^{2}-\omega
^{2}/c^{2}}$ \cite{next}. A generic mode of the dielectric is classified
according to its azimuthal, radial and vertical quantum numbers $(m,n,p)$.
If $m=0,$ the mode has cylindrical symmetry and can be either TE or TM.
Hybrid HEM modes arise whenever $m>0$. There is no TEM mode for a dielectric
guide. The agreement beween the calculated resonances and the data is found
to be within $2\%$ for all the peaks \cite{next}.

The quality factors $Q$ of the resonances are determined by the observed
widths $\gamma $ in the frequency domain, $Q=f/\gamma $. For a single
resonator, estimates of $Q$ are possible by calculating the ratio between
the energy stored per unit time and the average power dissipated \cite
{Kajfeh88}. Owing to the {\sl localized} nature of the field eigenmodes,
losses due to the open boundary conditions at the edge of the plates are
irrelevant, power dissipation being introduced by dielectric and conductor
losses. Detailed calculations, which yield results consistent with the
measurements, indicate that finite absorption in the metallic plates
outweighs dielectric losses by at least a factor 5, thus providing the
leading dissipation mechanism \cite{next}. In an equivalent time-domain
picture, this implies that the metal acts as an environmental decay channel
for the bound electromagnetic modes, the coupling between the dielectric and
the metal being proportional to the copper surface resistence.

When two dielectric resonators are placed {\sl in proximity }to each other,
each resonance splits into two. The doublets have the structure of {\sl a
broad resonance at a lower frequency }$f_{l}\ $ accompanied by {\sl a narrow
resonance at higher frequency }$f_{h}${\em .} This effect is particularly
pronounced for TM modes ($H_{z}=0$). The most noteworthy example is the 
TM$_{011}$ single-disk resonance found at $9.45$ GHz with 
$Q_{0}\approx 70$, which splits into two peaks with 
$Q_{h}\approx 2400$ and $Q_{l}\approx 50$ for an edge 
distance $d=1.0\,$mm (Fig. 2). The narrow
peak can be experimentally assigned to an {\sl antisymmetric }$E_{z}$-field
configuration by establishing electric field coupling with the pick-up
antenna and by probing the behavior at the mirror symmetry plane \cite{next}. 
The doublet splitting as a function of the disk separation $\Delta f(d)=
f_{h}-f_{l}$ is displayed in Fig. 3(a). The splitting vanishes exponentially
with $d$ until the limit of noninteracting resonators is approached. The
measured decay constant is in good agreement with the single-disk value $
\kappa _{r}=0.45$ mm$^{-1}$, as expected on the basis of semiclassical
estimates in a tunneling regime where $\kappa _{r}d\geq 1$ \cite{Creagh96}.
The widths $\gamma _{l}$, $\gamma _{h}$ and their ratio $\gamma _{l}/\gamma
_{h}$ are plotted in Fig. 3(b) and in Fig. 4 (Dots) respectively. Again,
single-disk behavior is recovered for sufficiently large $d$, where $\gamma
_{l},\gamma _{h}\rightarrow \gamma _{0}.$ For small separations, the width 
$\gamma _{h}$ is highly suppressed, leading to the high $Q$-values noted
above. As indicated by Fig. 4, a maximum ratio $\gamma _{l}/\gamma _{h}$ 
$\approx 50$ is seen at $d=0.73$ mm. It is very remarkable that, thanks
to the proximity effect, $Q^{\prime }s$ in the range of $10^{3}$ are
achievable without resorting to closed-walls cavities.

A quantitative account of the above results can be only achieved by
numerically solving Eq. (1) with the appropriate boundary conditions. Even
if the problem is simplified since the $z$-dependence is separable, an
accurate calculation of the lineshape factors $Q$ requires the complete
knowledge of the electric and magnetic field distribution within the cavity
volume. By referring to \cite{next} for more detail on the full
electromagnetic analysis, our primary goal here is to gain simple
qualitative insights. Let us focus henceforth on the TM$_{011}$
configuration. In the two-disk system, the splitting into modes of well
defined parity is easily understood as a consequence of the perturbation
introduced by the resonator-resonator coupling. The latter is known to take
contributions only from the unperturbed evanescent fields $E_{z}({\bf r})=
\cos (k_{z}z)\,E_{z}(x,y)$ of each resonator \cite{Kajfeh88}.
Experimentally, the antisymmetric mode is found to be able to store an extra
amount of electromagnetic energy compared to the symmetric one \cite{next}.
This takes place through an amplification of field components (e.g., $
E_{z},H_{z}$), which do not contribute to power dissipation, whereby the
higher observed $Q$.

The fact that the stabilisation of the antisymmetric mode only manifests at
small separations suggests to picture the phenomenon in terms of a {\sl 
collective }effect arising when two discrete states (the bound single-disk
modes) are coupled to each other and, in addition, to a {\sl common}
environment (the metal) that renders them unstable. Similar effects are
found in quantum physics, where they require an appropriate modification of
the standard Weisskopf-Wigner decay theory \cite{Barnett}. The
correspondence between electromagnetic ({\sc em}) and quantum mechanical (
{\sc qm}) systems is well established for stationary problems \cite
{Sridhar91a}. In particular, taking into account the boundary conditions at
the dielectric surface, the component $E_{z}(x,y{\bf )}$ in the waveguide
plays the role of the wavefunction $\psi $ in a 2D quantum mechanical
system, the dielectric medium corresponding to a square potential well at a
{\sl \ fixed }energy \cite{Heller98}. Accordingly, the two-disk system maps
into a 2D tunneling problem. In the presence of losses, the damping of the 
{\sc em} field amplitude is usually accounted through a {\sl complex}
frequency $f-i\gamma /2$ whose imaginary part provides the time decay rate 
\cite{Kajfeh88}. Despite the fact that, due to the different time-dependent
equations of motion, the correct mapping to complex energies of quantum
unstable states is a nonlinear relation of the form $(f-i\gamma /2)_{\text{
{\sc em}}}^{2}\Leftrightarrow (\varepsilon -i\gamma /2)_{\text{{\sc qm}}}$,
a {\sc qm} configuration which is stabilized against decay will still be
mapped into an {\sc em} non-decaying state.

A simple argument supporting the stability of the antisymmetric state goes
as follows. Let $\left| L\right\rangle ,\,\left| R\right\rangle $ denote
degenerate ket states localized in the left, right well respectively. The
two levels are coupled to each other by a tunneling perturbation of the form 
$H_{T}=-T\,[\left| L \right\rangle \left\langle R \right| +
\left| R \right\rangle \left\langle L \right| ]$, $T>0$, 
and to a common continuum of states with a strength $W_{L},W_{R}$. 
It is possible to show that the
coupling to the environment mediates an extra interaction between the two
discrete states, which strongly affects the decay properties of the combined
system and thereby its spectral response \cite{Viola}. If $W_{L}=W_{R}$, one
predicts that the symmetric combination $[\left| L\right\rangle +\left|
R\right\rangle ]$ corresponds to a Lorentzian resonance line at frequency 
$f_{S}$, whose width is twice larger than the width of each single level
coupled to a continuum of the same strength, while the antisymmetric
combination $[\left| L\right\rangle -\left| R\right\rangle ]$, found at
frequency $f_{A}$, is completely stabilized. This behavior can be regarded
as the counterpart of Heller's predictions for the proximity effect based on
a point scatterer model \cite{Heller96}. It is worth mentioning that the
overall effect of the extra interaction indicated above is to dress the
tunneling matrix element $T$ with an additional {\sl imaginary} part.

The actual situation, where the width of the antisymmetric mode is limited
by the dielectric losses, would be more adequately modeled by invoking two
distinct environments. To complement the previous analysis, we also explored
a phenomenological description of dissipation in terms of an effective
non-hermitian Hamiltonian \cite{Datta}. By introducing a drastic
approximation, we only consider an effective 1D potential projected along
the horizontal symmetry axis \cite{note}. Within the theory of
multidimensional tunneling, this is supported by the fact that the
interaction is dominated by instanton orbits between the two centers \cite
{Creagh96}. Thus, we use a potential energy function of the form $V_{\text{
pot}}(x)=V_{0}+iV_{1}$ (for $|x|>d/2+D$), $iV_{2}$ (for $d/2<|x|<d/2+D$), 
$V_{b}+iV_{1}$ (for $|x|<d/2$), all parameters being real numbers. The
imaginary terms $iV_{1},$ $iV_{2}$ account for the losses outside and inside
the double-well region respectively. We assume $\left| V_{1,2}\right|
/V_{0,b}\ll $ $1$.

The real part of $V_{\text{pot}}(x)$ is depicted in Fig. 4 (inset). We allow
for the possibility of a height barrier $V_{b}\neq V_{0}$ to effectively
include corrections arising from the 2D nature of the problem. The presence
of extra-contributions to the tunneling interaction, which are lost in the
1D model, is simulated by a more transparent barrier. The noninteracting
limit corresponds to $d\rightarrow \infty $. For finite $d$, even and odd
states are generated, with eigenenergies $E_{P}=\varepsilon _{P}-i\gamma
_{P}/2$, $P=S,\,A$. If $\kappa =\kappa _{r}+i\kappa _{i}$ denotes the
inter-well wave vector, each complex eigenvalue has a structure involving
exponentials $e^{-\kappa _{r}d}$, convoluted with oscillating functions of 
$\kappa _{i}d$ whose details depend on the state $\left| L\right\rangle $, 
$\left| R\right\rangle $ \cite{next}. The signature of tunneling shows up
through the exponential dependence of the energy splitting, $\Delta
\varepsilon =\varepsilon _{A}-\varepsilon _{S}\approx e^{-\kappa
_{r}d}F(\kappa _{i}d)$. The oscillatory terms in $F$ are responsible for a
``rippled'' structure of the splitting decay, which is apparent in the data
(Fig. 3(a)).

The transcendental equations determining $E_{A}$ and $E_{S}$ have been
solved numerically for different sets of parameters with both $V_{b}=V_{0}$
and $V_{b}<V_{0}$ and the results compared with the experimental ones \cite
{next}. The leading exponential decay of the energy separation and the
asymmetric small-distance splitting of the widths are correctly predicted.
However, we find a major difference between the two models in their
capability to reproduce the exceedingly large ratio between the widths. In
the simulations with $V_{b}=V_{0},$ we were unable to reach ratios larger
than 4, regardless of the values of the parameters $V_{1},V_{2},$ mainly
affecting the absolute range of the widths. This order of magnitude is in
agreement with independent results on symmetry splittings of resonances due
to semiclassical creeping orbits \cite{Wirzba96}. In the reduced-height
configuration, the ratio $\gamma _{s}/\gamma _{a}$ can be controlled over a
broad range (up to 50) by varying $V_{b}/V_{0}$. Some representative
behaviors are summarized in Fig. 4 for barrier opacity $V_{b}/V_{0}=1,$ 
$1/4, $ $1/9$. Maximum stability of the antisymmetric mode is reached at an
intermediate distance $d$, corresponding to the $\gamma _{s}/\gamma _{a}$
-peak value. The better qualitative agreement attainable with the
reduced-height model suggests that {\sl dimensionality }effects also play a
key role in the experiment.

In order to confirm the conclusion that a judicious use of symmetry leads to
a $Q$-sharpening, we carried out experiments on a 3-disk system, with the
disks placed at the vertices of an equilateral triangle. Resonance triplets
are observed. In analogy to the 2-disk system, we predict that modes
tranforming antisymmetrically with respect to reflections in each of the
vertical mirror planes show enhanced stability against dissipation \cite
{next}. We focus on a single-resonator mode with $m=3$ at 10.8 GHz, whose
behavior is displayed in Fig. 5. A very sharp component at intermediate
frequency is clearly seen. The $Q$ factor is increased by roughly a factor
20 compared to the original one. The sharpening effect turns out to be very
sensitive against symmetry-breaking effects. The influence of a geometric
symmetry-breaking has been studied by shifting one of the disks by $b$ along
the bisectrix of the triangle. It is evident from Fig. 5 (inset) that the
sharp resonance is dramatically affected, the $Q$ factor being exponentially
degraded. No sensible change is observed for any of the broad resonances.

In the spirit of the original definition by Heller \cite{Heller96}, our
observations indicate that interesting proximity phenomena arise in the
spectral response of nearby systems. Proximity resonances have been recently
detected in the scattering of a TEM\ electromagnetic mode in a
parallel-plate waveguide \cite{Heller98}. Despite some superficial
similarity with the present work, it is essential to realize that our
experiments probed a completely different regime of the microwave field,
where direct evanescent-wave coupling between {\sl bound }modes rather than 
{\sl scattering} resonances from two dielectrics illuminated by the same
wave field were investigated. In particular, the power law behavior
characterizing scattering states \cite{Heller96} should be contrasted with
the exponential dependences that are intrinsically associated with
tunneling. Thus, bound-state proximity resonances form a novel complementary
manifestation of a similar physical phenomenon, whose detailed understanding
poses new challenges to both numerical simulations and semi-classical
treatments.

Our results have a variety of implications. First, the tunneling interaction
in a 2D integrable potential has been probed sensitively. There is no
difficulty, in principle, to extend the experimental work to chaotic
potentials where important results on complex periodic orbits theory have
been obtained \cite{Creagh96}. Second, the experiments demonstrate how
symmetry properties can be usefully exploited to protect a system against
the effects of its environment. This general mechanism provides a unifying
explanation for proximity resonances, regardless of the unbound (scattering)
or bound (confined) nature of the wave field. Third, our work can be related
to recent observations of the symmetry splitting between optical modes in
photonic molecules \cite{Bayer98} where, however, the behavior of widths was
not addressed. It is conceivable that some counterpart of proximity
phenomena may be relevant on the mesoscopic scale as well. From the
practical perspective, the possibility of symmetry-based $Q$-amplification
in electromagnetic or photonic structures represents another exciting area
of applications. Finally, the electromagnetic phenomenon evidenced here
displays intriguing similarities with concepts investigated in the context
of quantum dissipative processes \cite{Barnett}, \cite{Lidar98}. The
possibility of establishing some mapping between the electromagnetic and
quantum realm even in the presence of dissipative mechanisms would clearly
open up a fruitful arena of interchange and deserves further investigation.

Work at Northeastern was supported by NSF-PHY-9752688. We thank V. Kidambi
for help with the data analysis software and N. D. Whelan for discussions.


\newpage

\begin{figure}[h]
\mbox{ \epsfig{width=.8 \linewidth,file=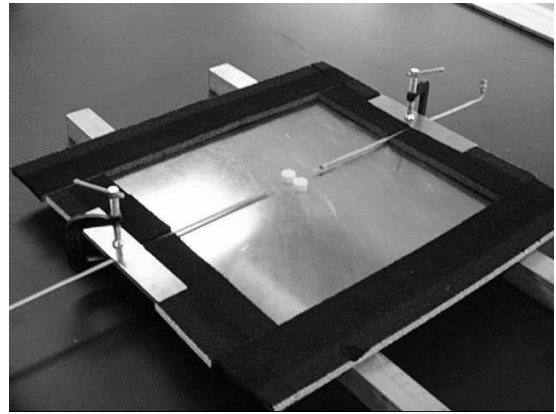} }
\vspace*{6mm}
\caption{ View of the experimental setup. Shown in the photo are two
cylindrical dielectric resonators, and input and output antennas. }
\label{Fig1}
\end{figure}

\vspace*{-7mm}

\begin{figure}[h]
\mbox{ \epsfig{width=8.5cm,file=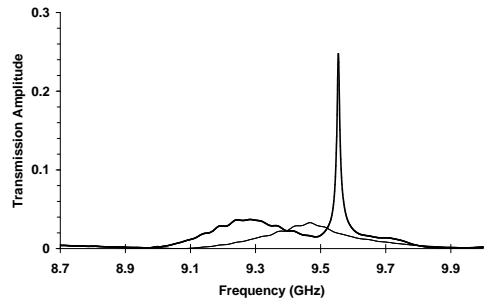} }
\vspace*{-.4cm}
\caption{ Experimental proximity resonances for two coupled resonators
operating in the TM$\mbox{}_{011}$ mode at 9.45 GHz. }
\label{Fig2}
\end{figure}

\begin{figure}
\mbox{\epsfig{width=8.5cm,file=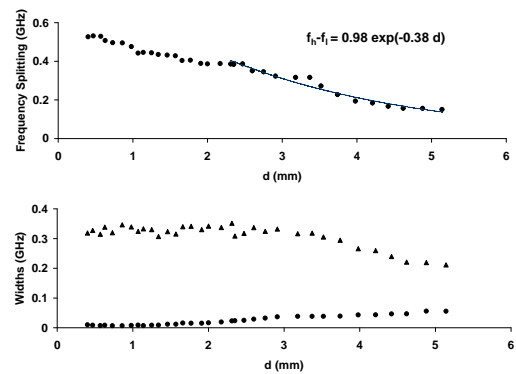} }
\vspace*{-.6cm}
\caption{(Top) Doublet frequency splitting $\Delta f$ and (Bottom) Widths 
$\gamma_l,\gamma_h$ vs. distance $d$ for the two-disc proximity resonance
around 9.45 GHz. The result of an exponential fit in the region $\kappa_r d
\geq 1$ is also shown. }
\label{Fig3}
\end{figure}

\newpage
\begin{figure}
\mbox{\epsfig{width=.95 \linewidth,file=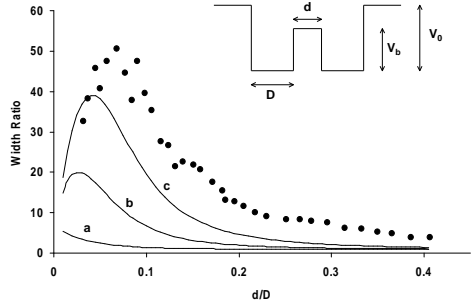}}
\vspace*{-.4cm}
\caption{Ratio between symmetric and antisymmetric width vs. distance $d/D$
for experimental data (Dots) and various implementations of the double-well
potential shown in the inset. Units where $\hbar=2m=1$ have been chosen. 
$V_0 =900$ in units $D^{-2}$ and $V_1/V_0=-3\cdot 10^{-2}$, $V_2/V_0=-3\cdot
10^{-6}$. The barrier height is $V_b/V_0$=1 (a), 1/4 (b), 1/9 (c). }
\label{Fig4}
\end{figure}

\begin{figure}
\mbox{\epsfig{width=.95 \linewidth,file=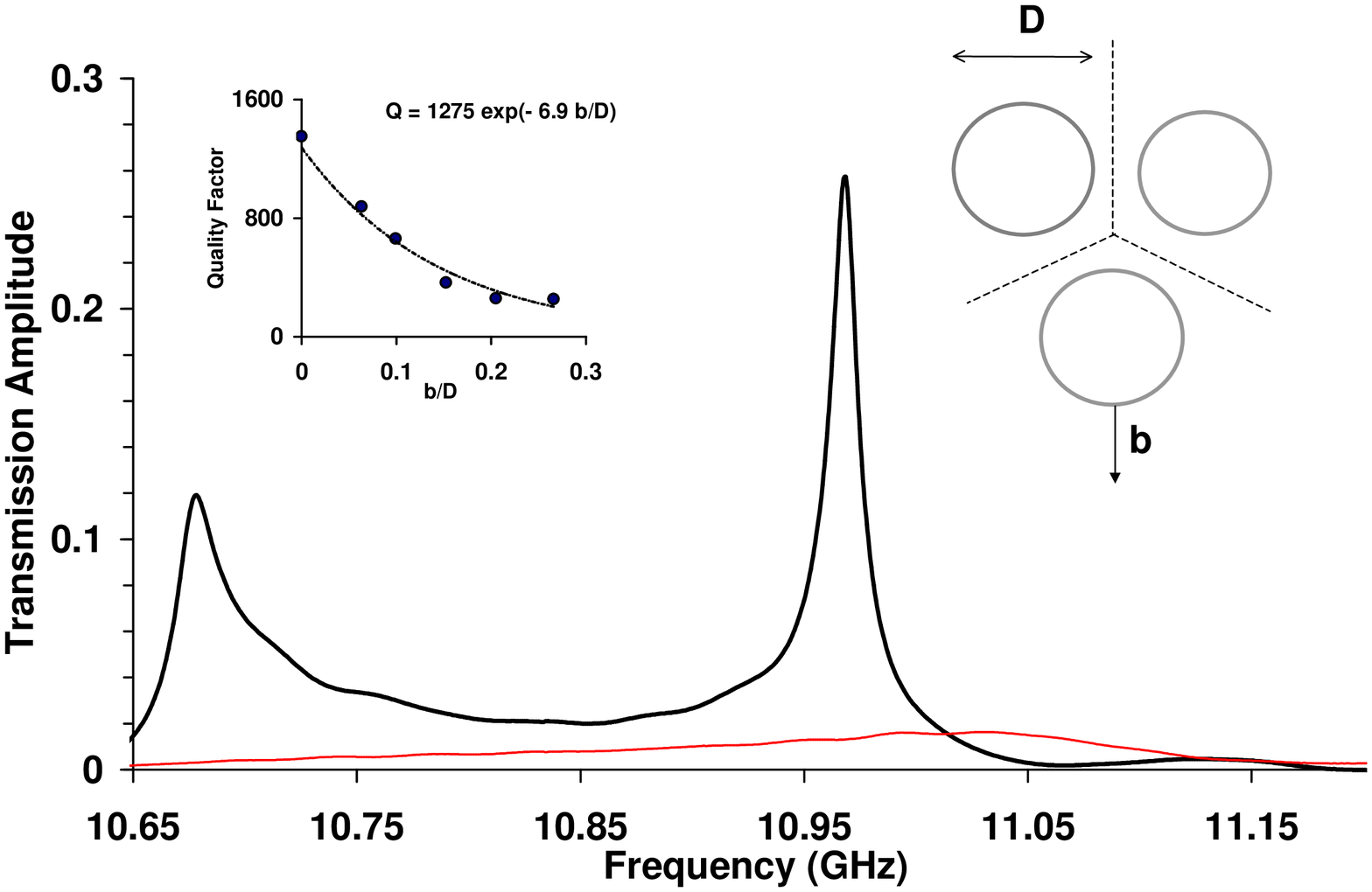}}
\vspace*{-.4cm}
\caption{ Resonance triplet for three coupled resonators operating in the
hybrid (3,1,1) mode at 10.8 GHz. (Inset) Quality factor of the sharp
component as a function of the symmetry-breaking parameter $b/D$. }
\label{Fig5}
\end{figure}

\end{document}